\documentclass[%
letterpaper,%
amsfonts,amssymb,amsmath,%
notitlepage%
]{revtex4-1}

\usepackage[breaklinks]{hyperref}
\usepackage{graphicx}
\usepackage{amsmath,amssymb}

\newcommand{\e}[1]{\mathrm{e}^{#1}}
\newcommand{\kB}{k_\mathrm{B}}
\renewcommand{\d}{\mathrm{d}}

\begin{document}

\title{Sampling in the multicanonical ensemble: Small He clusters in W}

\author{Thomas Vogel}
\email[E-mail: ]{tvogel@lanl.gov}
\author{Danny Perez}

\affiliation{Theoretical Division (T-1), Los Alamos National Laboratory, Los Alamos, NM 87545, USA}

\begin{abstract}

  We carry out generalized-ensemble molecular dynamics simulations of
  the formation of small Helium (He) clusters in bulk Tungsten (W), a
  process of practical relevance for fusion energy production. We
  calculate formation free energies of small Helium clusters at
  temperatures up to the melting point of W, encompassing the whole
  range of interest for fusion-energy production.  From this,
  parameters like cluster break-up or formation rates can be
  calculated, which help to refine models of microstructure evolution
  in He-irradiated Tungsten.

\end{abstract}

\date{April 3, 2014}
\maketitle

\section{Introduction}
\label{sec:Intro}

Many challenges have to be addressed to enable energy production from
nuclear fusion. While issues related to plasma stability are
considerable, it is increasingly clear that material stability in the
extreme conditions typical of the operation of such reactor are
critical. For example, tungsten is a leading candidate for the
construction of plasma-facing
components~\cite{Janeschitz2001,Bolt2004}, due to comparatively
favorable properties such as very high melting point, low sputtering
yield, etc. One serious problem however is that the irradiation of W
by Helium ions incoming from the plasma can cause serious
modifications in the microstructure of the exposed
materials~\cite{Nishijima2004,Tokunaga2005,Takamura2006,Baldwin2008,Baldwin2009}.
In particular, He bubbles nucleating from small He clusters inside the
material can grow, coalesce, and burst, severely damaging the surface.
Since small helium clusters (of sizes $N\simeq7$) serve as nuclei of
such bubbles, it is of great technological interest to investigate the
behavior of such clusters in W~\cite{Perez2014,Wirth2014}.

In this paper we focus on one key detail of such a study, the
determination of the formation free energy of He clusters. We perform
molecular dynamics (MD) simulations in the multicanonical
ensemble~\cite{Hansmann1996,Junghans2014} and measure the
distributions of He cluster compositions. We obtain the simulation
weights through STMD (Statistical Temperature
MD)~\cite{Kim2006,*Kim2007}, a molecular dynamics
protocol derived from the Wang--Landau
approach~\cite{Wang2001a,*Wang2001b}.

\section{Simulation protocol}

\subsection{Theory review and estimation of simulation weights}

The multicanonical (muca) ensemble~\cite{Berg1991,*Berg1992} is the
ensemble where the distribution of potential energies (or any other
reaction coordinate/order parameter) becomes flat. In
practice, the simulation weights $w_{muca}(E)$ (which would be
the Boltzmann factors for the canonical ensemble) required to sample from this
ensemble, be it using Monte Carlo (MC) or molecular dynamics (MD)
methods, are not known {\em a priori} and have to determined first. Since
$w_{muca}(E)$ is related to the density of states $g(E)$ via
the ensemble-defining condition
\begin{equation}
  \label{eq:mucaP}
  P_{muca}(E)\propto g(E)\,w_{muca}(E)\stackrel{!}{=}\mathrm{const.},
\end{equation}
one can obtain thermodynamic averages at \emph{any}
temperatures by reweighting measured data with respect to the
simulation weights, as will be shown below. 

For a muca MD simulation one can, in principle, use any canonical
integrator at a reference temperature $T_0$ using an effective
potential $V_{muca}$ leading to the flat, multicanonical distribution
instead of the usual, raw potential (which would lead to a canonical distribution).
Introducing the entropy $S(E)=\kB\ln g(E)$ (where
$\kB$ is the Boltzmann constant), one defines from
Eq.~(\ref{eq:mucaP}) the effective potential $V_{muca}(E)$:
\begin{equation}
  w_{muca}(E)\propto \e{-\kB^{-1}S(E)}=:\e{-V_{muca}(E)/(\kB T_0)}\,,\quad
  \mathrm{and}\quad V_{muca}(E)=T_0\,S(E)\,.
\end{equation}
Interatomic forces are then calculated via the gradient of the
effective potential with respect to the particle coordinates:
\begin{equation}
  \label{eq:mucaf}
  f_i^{muca}=-\frac{\d V_{muca}(S(E(q_1,\ldots,q_{3n})))}{\d q_i}
  =-T_0\,\frac{\partial S(E)}{\partial E}\,\frac{\d E(q_1,\ldots,q_{3n})}{\d q_i}\,.
\end{equation}
Since the last term equals minus the canonical (conventional) forces $f_i$ and the second
one defines a temperature via the thermodynamic relation
$T(E)^{-1}:=\partial S(E)/\partial E$, we can write the multicanonical
forces simply as rescaled canonical forces:
\begin{equation}
  \label{eq:mucaf2}
  f_i^{muca}=\frac{T_0}{T(E)}\,f_i\,.
\end{equation}
The function $T(E)$ is still unknown and estimating it is equivalent
to estimating $g(E)$, $V_{muca}(E)$, or $S(E)$. 
Following the Wang--Landau (WL) scheme for MC simulations, Kim
\textit{et al.}  proposed a method (STMD~\cite{Kim2006,*Kim2007}) to
estimate $T(E)$ during a MD simulation. Once the estimator
$T^\prime(E,t)$ has converged, $S(E)$ can be estimated by direct
numerical integration. In STMD, one would start (at simulation time
$t=0$) with a constant initial guess $T^\prime(E,t=0)=T_{init}$ (which
is equivalent to a canonical MD simulation at $T_{init}$) and update
$T^\prime(E,t+\Delta t)$ via
\begin{equation}
  \label{eq:STMD_Tupdate}
  T^\prime(E_{{act}\pm1},t+\Delta t)
  =\frac{T^\prime(E_{{act}\pm1},t)}{1\mp\delta_\beta\,T^\prime(E_{{act}\pm1},t)}\,,
\end{equation}
with $\delta_\beta:=\kB \ln f_{\mathrm{WL}}/2\Delta E$. We assume that
$T^\prime(E,t)$ was binned (with $\Delta E$ being the energy bin
width) and the underlying WL procedure is just the update of the single
bin containing the current energy $E_{act}$ via $\ln g^\prime(E_{act},t+\Delta
t)=\ln g^\prime(E_{act},t)+\ln f_{\mathrm{WL}}$. $f_{\mathrm{WL}}$ is usually
called the modification factor and decreases during the simulation as
in the WL scheme~\cite{Wang2001a,*Wang2001b}, $y^\prime(x)$ always refers to an
estimator for the true function $y(x)$. The derivation of
Eq.~(\ref{eq:STMD_Tupdate}) is straightforward and details can be found
in the original publications~\cite{Kim2006,*Kim2007}. It can be noted 
that this scheme leads to exactly the same dynamics as
obtained in a metadynamics simulation if $T^\prime(E,t)$ is
updated on the basis of Gaussian kernel functions, provided all method
parameters are chosen consistently~\cite{Junghans2014}.

\begin{figure}[b]
  \includegraphics[width=\textwidth]{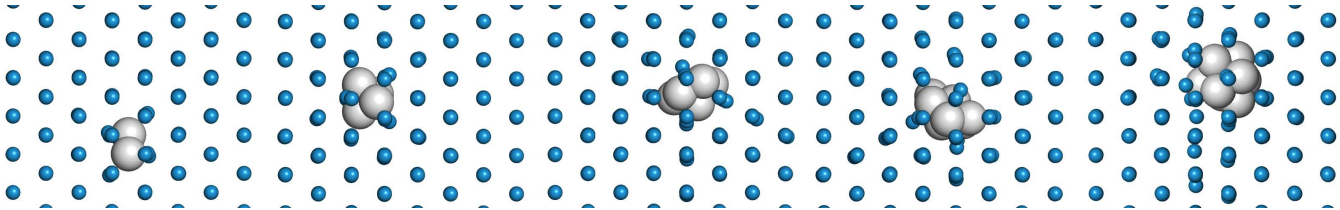}
  \caption{Ground-state structures for $N=2$, $4$, $6$, $8$, and $10$
    (from left to right) He atoms in bulk W. While the cluster of two
    He atoms (left) only slightly and locally affects the W structure,
    larger clusters also disturb next-neighbor positions and
    eventually eject W interstitial crowdions, leading to the
    nucleation of a nano-bubble (right).}
  \label{fig:gs}
\end{figure}

\subsection{Physical system, simulation details, and measurements}

The systems we are simulating consists of $432$ Tungsten
($6\times6\times6$ W unit cells) and $N$ Helium (He) atoms, which
interact via embedded atom potentials (W--W interactions:
\cite{Ackland1987}, modified in~\cite{Juslin2013}; He--He
interactions:~\cite{Beck1986a,*Beck1986b}, modified
in~\cite{Morishita2003}; He--W interactions:~\cite{Juslin2013}). For
illustration, Fig.~\ref{fig:gs} shows ground-state configurations for
$N=2$, $4$, $6$, $8$, and $10$, also showing the effect of small He
clusters on the structure of the surrounding W. Technically, we
perform simulations for $2\leq N\leq 8$ in the temperature range
$T_{min}=100\,\mathrm{K}\leq T(E)\leq T_{max}=3700\,\mathrm{K}$.
Outside this range, we effectively perform canonical simulations at
$T_{min}$ and $T_{max}$, respectively, by holding $T(E)$ fixed at
those values (cp.~\cite{Kim2006,*Kim2007}). However, it is almost
impossible to obtain reliable data regarding free-energy differences
for very low temperatures anyway since the probability of observing
free He atoms vanishes (see below). Therefore, in practice, we require
the histogram to be flat only for $T>300\,\mathrm{K}$ in order to save
time. The reference temperature $T_0=2200\,\mathrm{K}$ is imposed
using a Langevin thermostat and the particle positions evolve
following the stochastic Velocity-Verlet integration
scheme~\cite{melchionna07jcp}. The simulation box has a linear size of
$18.991$\,\AA, i.e., we simulate at constant volume, and periodic
boundary conditions are applied in all three directions.

\begin{figure}[t]
  \includegraphics[width=.32\textwidth]{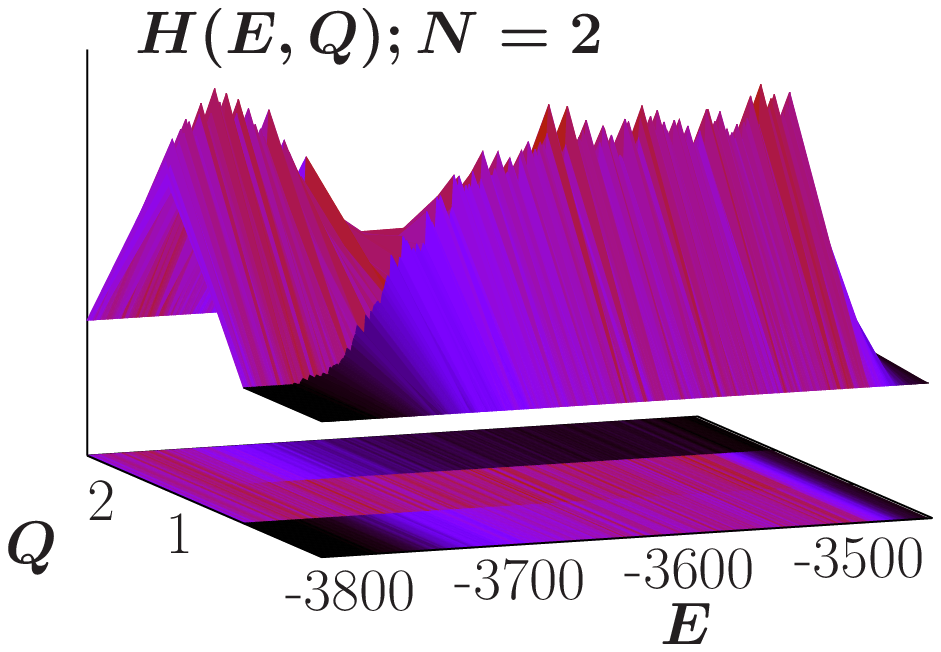}\hfill
  \includegraphics[width=.32\textwidth]{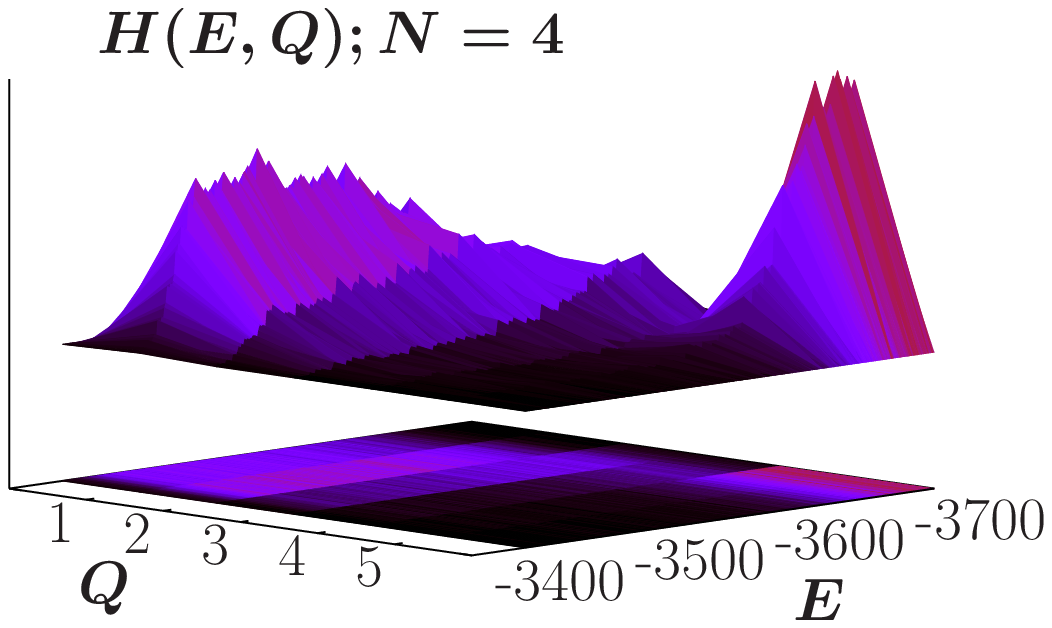}\hfill
  \includegraphics[width=.32\textwidth]{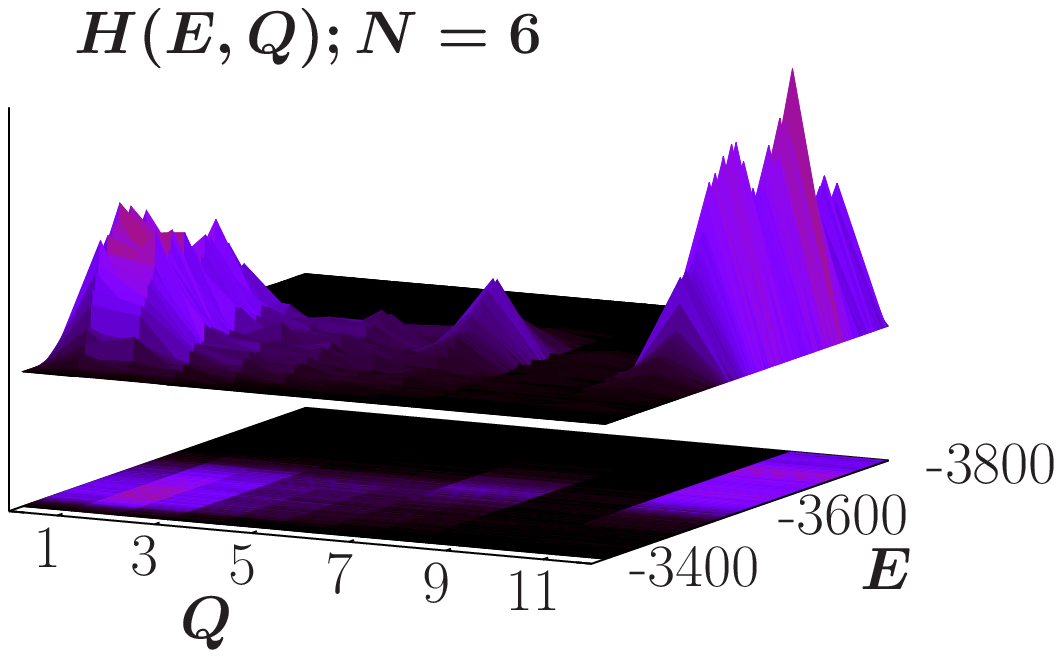}
  \caption{Examples of two-dimensional histograms of energy ($E$) and cluster
    composition ($Q$) as measured during multicanonical production
    runs.}
  \label{fig:raw_hist}
\end{figure}

After having estimated $T(E)$ via STMD, we fix it and perform muca MD
simulations based on Eq.~(\ref{eq:mucaf2}). During these production runs, we
measure the 2-dimensional histograms $H(E,Q)$, where $Q$ is the
He-cluster composition. In practice, the values of $Q$ are just
integers uniquely identifying every possible arrangement of the $N$ He
atoms into clusters (including the trivial cluster of size 1). For
$N=2$, there exist $c(N)=2$ such compositions: two single He atoms
(o-o) or one cluster of two (oo). For $N=3$, there are $c(N)=3$
compositions (o-o-o; oo-o; ooo), $c(4)=5$ (o-o-o-o; o-o-oo; o-ooo;
oo-oo; oooo), $c(5)=7$, $c(6)=11$, and so on. For large $N$, the
number of compositions $c(\sqrt{N})$ grows
exponentially~\cite{Kindt2013jctc}.  An off-lattice version of the
Hoshen--Kopelman algorithm~\cite{Hoshen1976} was implemented to
uniquely identify the cluster distribution and measure $Q$.
Figure~\ref{fig:raw_hist} shows three examples of histograms $H(E,Q)$
measured in the production runs.  We perform many independent runs for
each value of $N$ and use the combined histograms for further
calculations. All simulations ran on single CPUs, a small cluster
machine (less than 100 CPUs) was used to perform independent
production runs in parallel. Estimation of $T(E)$ took about one week
for each $N$, production ran for a few days.

\section{Results}

\begin{figure}[t]
\centering
\begin{minipage}[b]{.8\textwidth}
  \includegraphics[width=.48\textwidth]{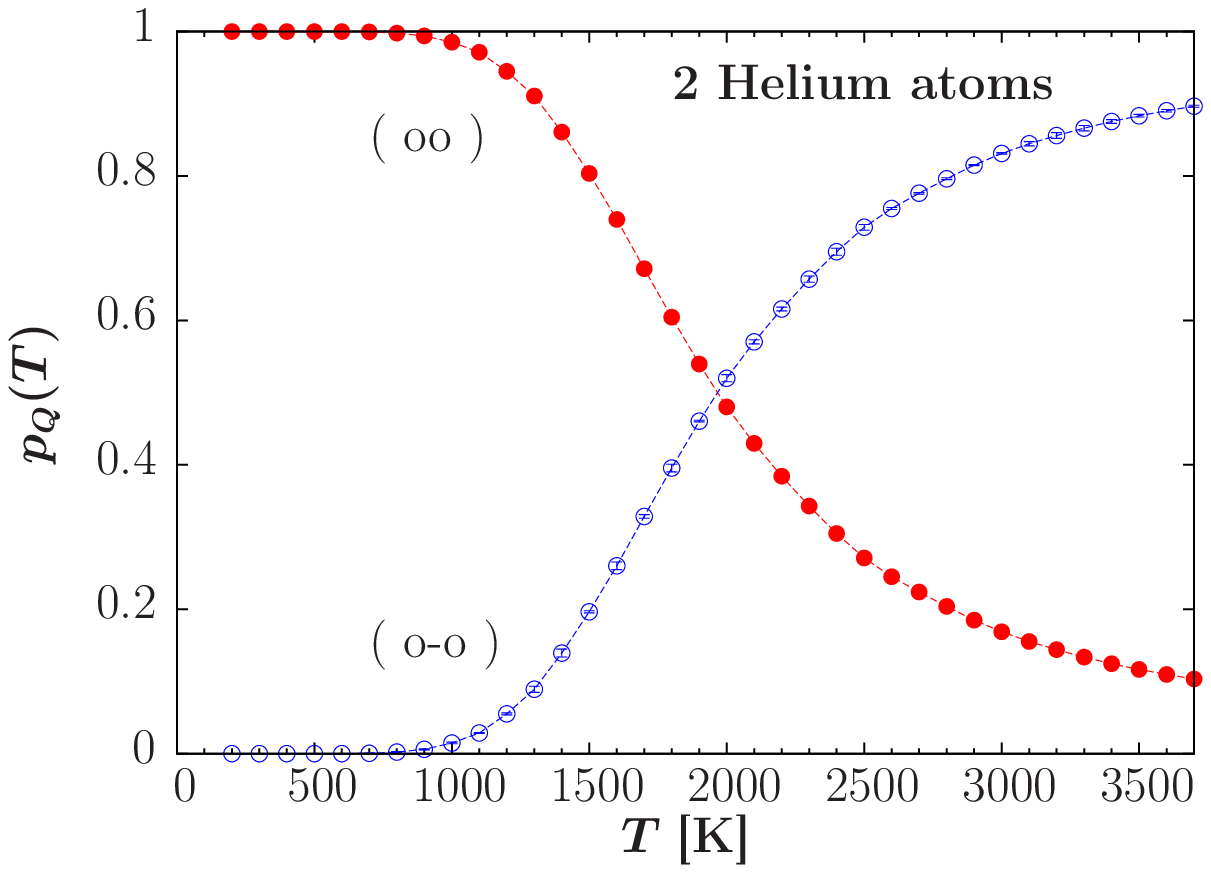}\hfill
  \includegraphics[width=.48\textwidth]{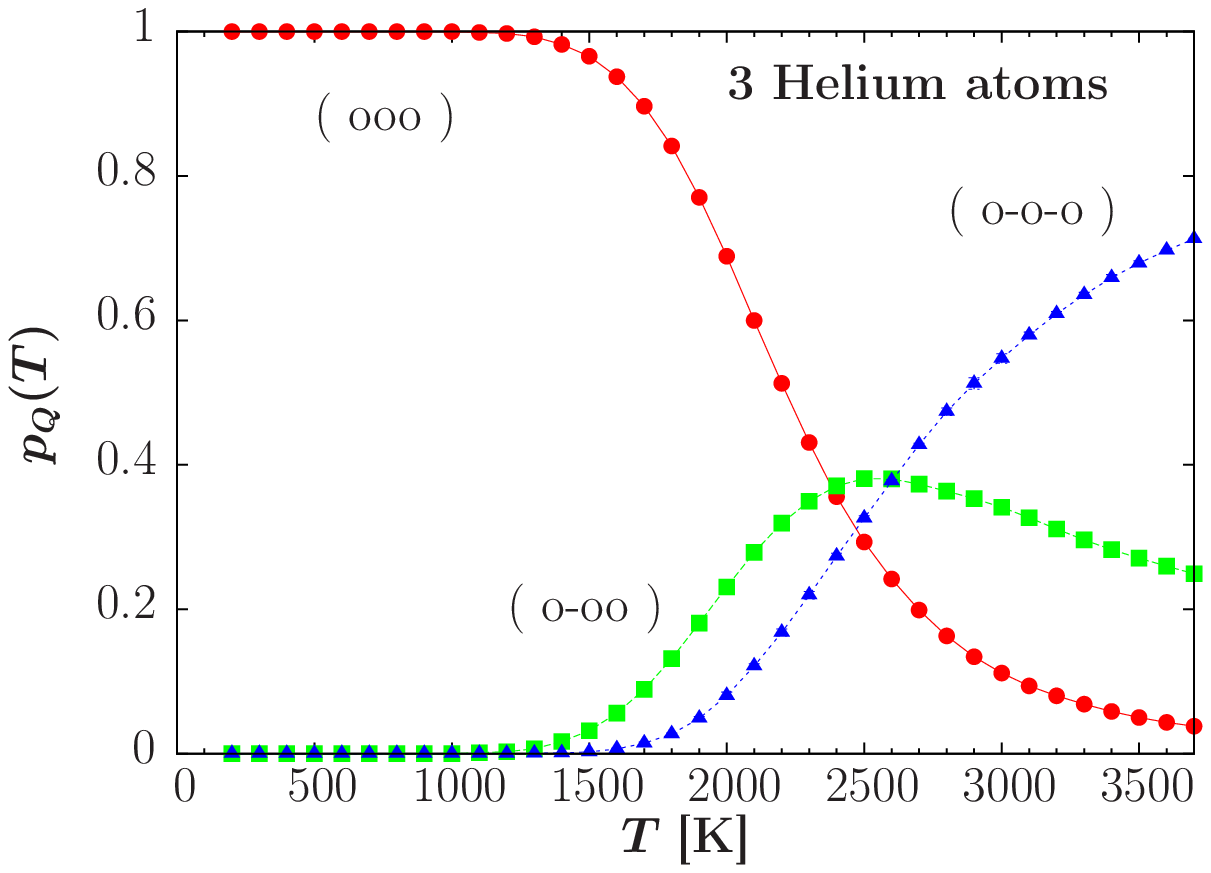}\\
  \includegraphics[width=.48\textwidth]{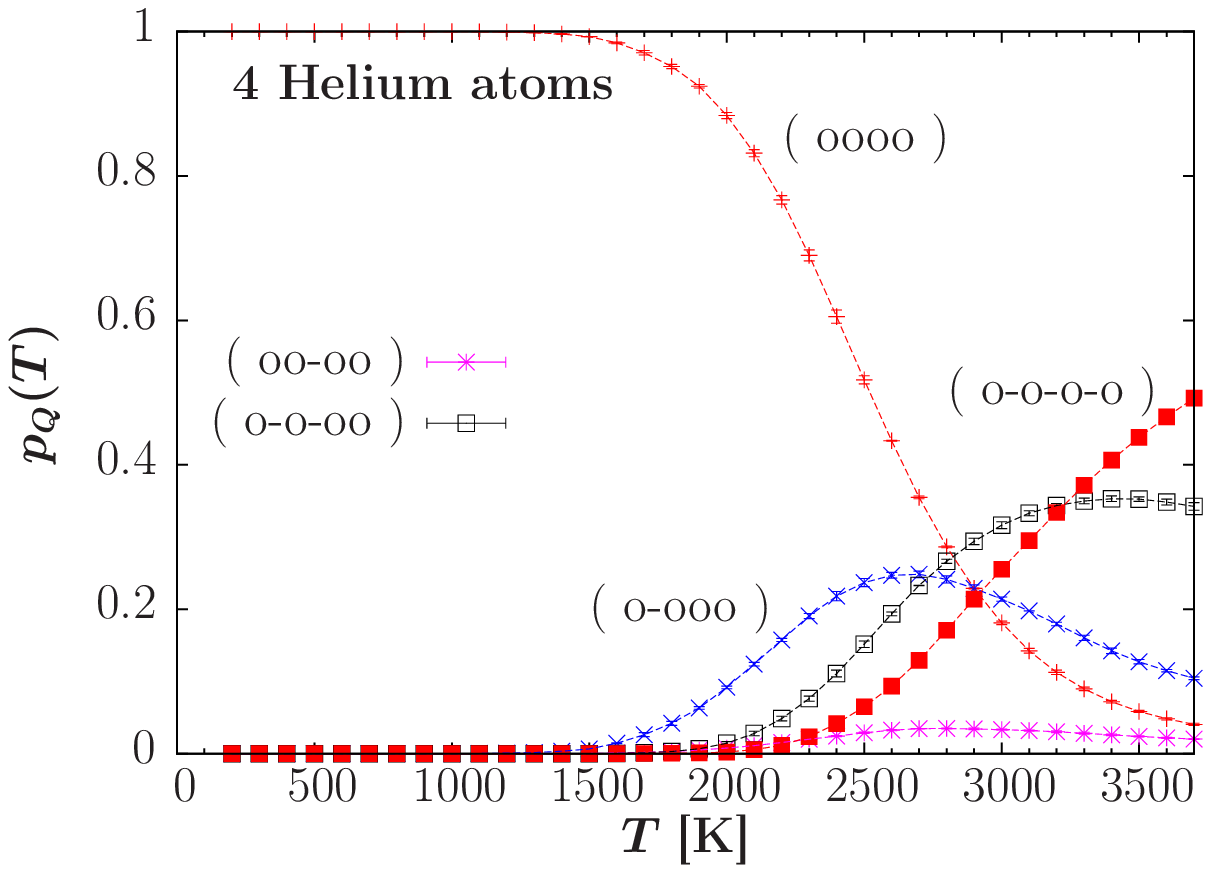}\hfill
  \includegraphics[width=.48\textwidth]{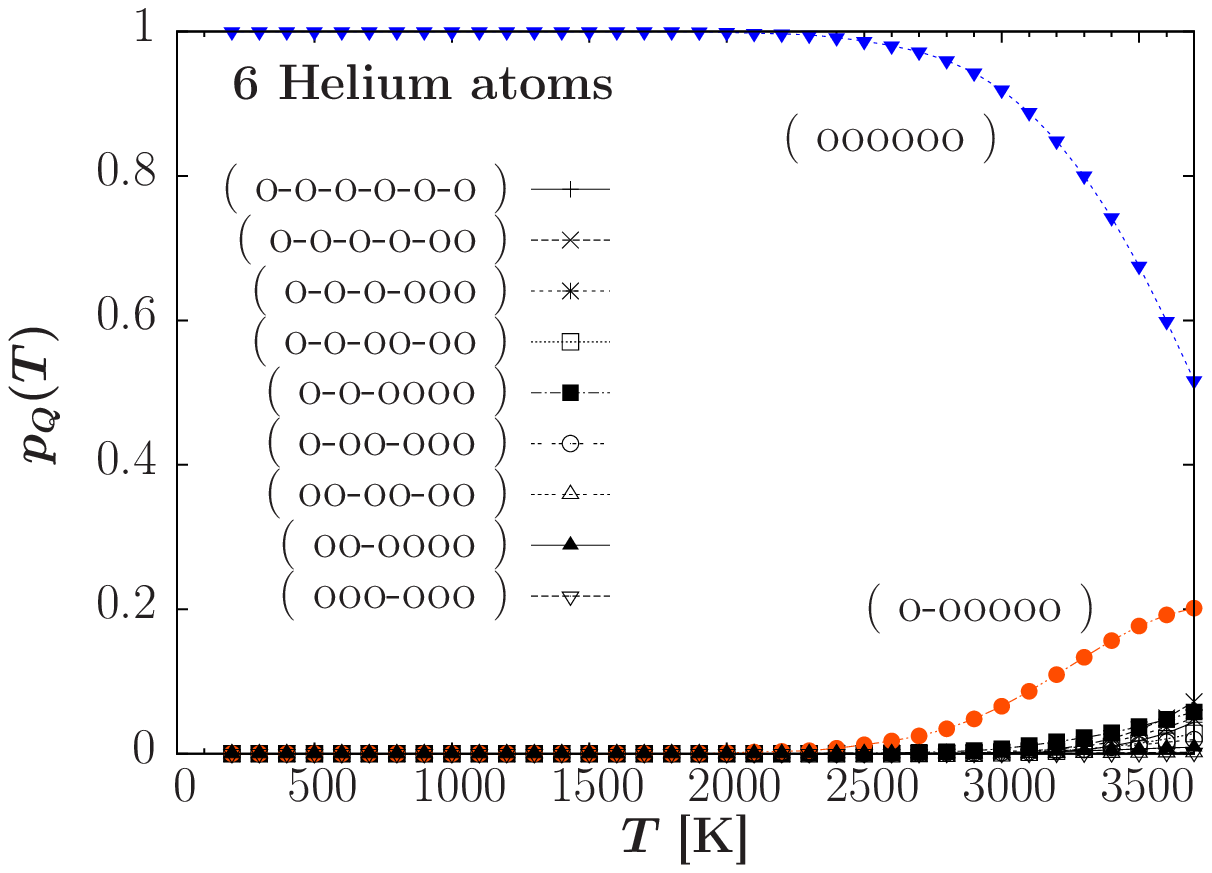}
\end{minipage}
\caption{Probabilities $p_Q(T)$ for cluster compositions to occur at
    different temperatures. Error bars are shown for all data, but
    might be smaller than the symbols.}
  \label{fig:probs}
\end{figure}

Canonical distributions of the cluster compositions $P_T^{can}(Q)$ for
all temperatures $T$ are obtained by reweighting, i.e., by dividing by
the applied simulation weights, multiplying with the Boltzmann
weights, and summing over all energies:
\begin{equation}
  P_T^{can}(E,Q)=w^{-1}_{muca}\,H(Q,E)\,\e{-E/\kB T}\,,\quad
  P_T^{can}(Q)=\sum_E P_T^{can}(E,Q)\,,
\end{equation}
where we use a sum instead of an integral because of the discrete
nature of the energy bins. With these distributions in hand, it is straightforward to
calculate the probabilities $p_Q(T)$ for certain cluster compositions
to occur at a given temperature~$T$. We plot $p_Q(T)$ for $N=2$, $3$,
$4$, and $6$ in Fig.~\ref{fig:probs}. The high-temperature boundary in
the plots corresponds to the melting temperature of Tungsten. For
$N=2$, we see that for $T<1000\,\mathrm{K}$ isolated He
atoms are basically never observed, while at $T\approx 2000\,\mathrm{K}$ we find single atoms and
cluster with about the same probability. Furthermore, it appears
that He cluster become more stable as $N$ increases. For
example, single He atoms start to split off a cluster of size $N=4$ at
$T\approx 1500\,\mathrm{K}$. For $N=6$, single He atoms
are basically never found at any temperature $T\lesssim 2500\,\mathrm{K}$. Even
for the highest temperatures, the only relevant compositions are
(oooooo) and (o-ooooo). However, note that these probabilities depend
on the volume of the simulation cell and might be different for
constant pressure simulations, for~example.
\begin{figure}[t]
\centering
\begin{minipage}[b]{.8\textwidth}
  \includegraphics[width=.48\textwidth]{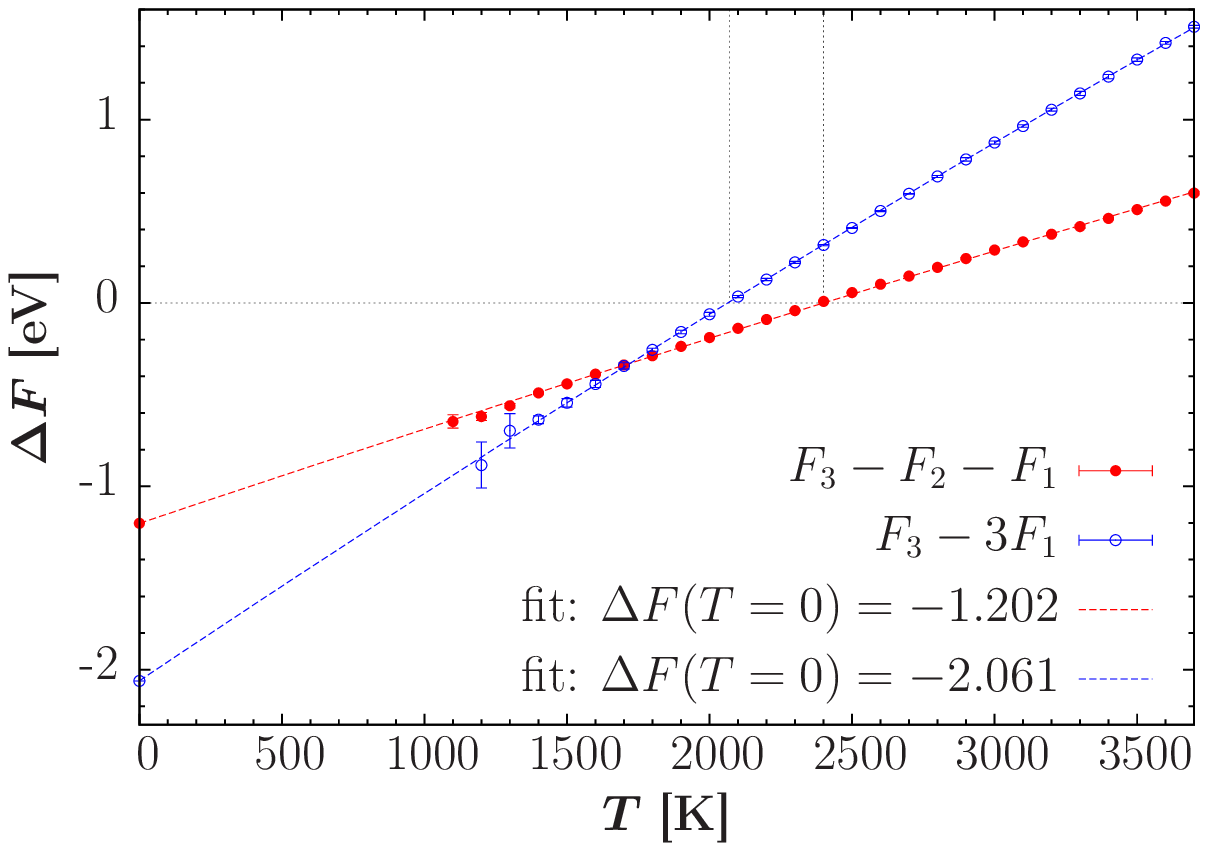}\hfill
  \includegraphics[width=.48\textwidth]{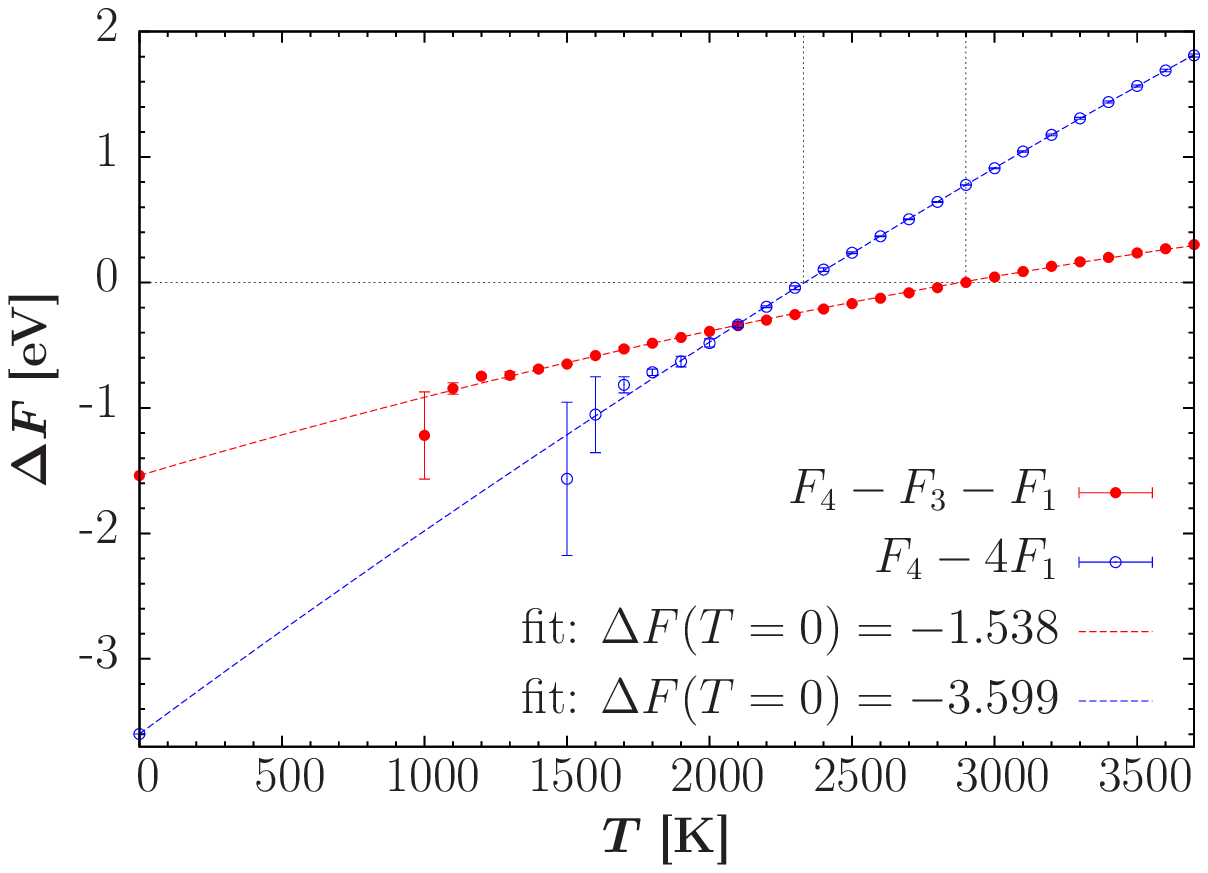}
\end{minipage}
  \caption{Formation free energy differences for different
    temperatures calculated from data shown in Fig.~\ref{fig:probs}.
    Left: 3 He atoms, right: 4 He atoms.}
  \label{fig:Fs}
\end{figure}

The probabilities $p_Q(T)$ provide a means to calculate free energy
differences. Following Kindt~\cite{Kindt2013jctc}, we write the free energy of
a He cluster of size $s$ as $F_s(T)=-k_\textrm{B}T\ln q_s$, where
$q_s$ is the partition function of that cluster (in particular, $q_1$
is the partition function of a free He atom). 
The free energy difference for the complete breakup of a cluster of
size $s$ into $s$ single atoms, for example, then reads:
\begin{equation}
  \label{eq:F1}
  F_s-sF_1=-k_\textrm{B}T\ln\left({q_s}/{q_1^s}\right)\,.
\end{equation}
Analogously, one defines free energy changes corresponding to the
split-off of a single atom from a cluster of size $s$ as:
\begin{equation}
  \label{eq:F2}
  F_s-F_{(s-1)}-F_1=-k_\textrm{B}T\ln\left({q_s}/{q_{(s-1)}q_1}\right)\,.
\end{equation}
The ensemble-averaged numbers ($\langle n_s \rangle$) of clusters of
size $s$ are functions of the partition functions
$q_s$~\cite{Kindt2013jctc}, hence we can write the free energy
differences as functions of the $\langle n_s \rangle$ (calculations
not shown). Since the $\langle n_s \rangle$ are related to the
probabilities $p_Q(T)$, ratios of cluster
partition functions are related to ratios of the probabilities
to observe certain cluster compositions. For example:
\begin{equation}
  \label{eq:ratios}
  \frac{q_2}{q_1^2}=\frac{p_{\textrm{(oo)}}}{2!\,p_{\textrm{(o-o)}}}\,,\quad
  \frac{q_3}{q_2q_1}=\frac{p_{\textrm{(ooo)}}}{p_{\textrm{(o-oo)}}}\,,\quad
  \frac{q_3}{q_1^3}=\frac{p_{\textrm{(ooo)}}}{3!\,p_{\textrm{(o-o-o)}}}\,,
\end{equation}
and so on. We show the formation free energies corresponding to
Eqs.~(\ref{eq:F1}) and~(\ref{eq:F2}) for $N=3$ and $4$ in
Fig.~\ref{fig:Fs} as illustrations. For low temperatures, we obtain relatively
large errors bars, the origin of which becomes clear when looking at
Eqs.~(\ref{eq:ratios}) and the data in Fig.~\ref{fig:probs}. For $N=4$
and $T=1600\,\mathrm{K}$, for example, the composition (o-o-o-o) is
extremely rare, the ratio $p_{\textrm{(oooo)}}/p_{\textrm{(o-o-o-o)}}$
being larger than $10^6$. For even lower temperatures, it is not
practical to calculate reliable free energy differences in this way.
However, at $T=0$, the free energy difference
reduces to the difference of the potential energies of the ground
state structures. This point can be used to obtain a fit to the free
energy over the whole temperature range. 

\section{Outlook}

Using differences in formation free energies one can
calculate cluster break-up or formation rates, and estimate effective
capture radii. The calculated cluster
formation rates, for example, are shown in~\cite{Perez2014} and were
validated by independent measurements from canonical molecular
dynamics simulations at fixed temperatures. These quantities are
essential to parameterize higher level models of microstructure evolution in
low energy He irradiated Tungsten~\cite{Wirth2014}, and to understand
and optimize the behavior of W in the extreme conditions of relevance
to fusion energy production.

\begin{acknowledgments}\noindent
Funding for this research was provided by Los Alamos National
Laboratory's (LANL) Laboratory Directed Research and Development ER
program. LANL is operated by Los Alamos National Security, LLC, for
the National Nuclear Security Administration of the U.S. DOE under
Contract DE-AC52-06NA25396. LA-UR-14-21924 assigned.
\end{acknowledgments}

\end{document}